\documentclass{phc-proc4-auth}

\usepackage{graphicx}
\usepackage{bm}
\usepackage{amssymb}

\begin{document}
\begin{frontmatter}

\title{Field-induced thermal metal-to-insulator transition in underdoped
LSCO}

\author{D.G.~Hawthorn},
\author{ R.W.~Hill},
\author{F.~Ronning},
\author{Mike~Sutherland},
\author{Etienne~Boaknin},
\author{M.A.~Tanatar\thanksref{MAT}},
\author{Johnpierre~Paglione},
\author{S.~Wakimoto},
\author{H.~Zhang},
\author{Louis~Taillefer\corauthref{cor1}\thanksref{LT}}

\address{Department of Physics, University of Toronto, Toronto, Ontario,
Canada M5S 1A7}
 \corauth[cor1]{Corresponding author: Louis.Taillefer@physique.usherb.ca}
\thanks[MAT]{Permanent address: Inst. Surf. Chem., N.A.S. Ukraine.}
\thanks[LT]{Current address: Department of Physics, University of
Sherbrooke, Sherbrooke, Quebec, Canada J1K 2R1.}

\date{\today}

\begin{abstract}
The transport of heat and charge in cuprates was measured in undoped and heavily-underdoped single crystal La$_{2-x}$Sr$_x$CuO$_{4+\delta }$ (LSCO).
In underdoped LSCO, the thermal conductivity is found to decrease with increasing magnetic
field in the $T \rightarrow 0$ limit,
in striking contrast to the increase observed in all superconductors, including cuprates at higher doping.
The suppression of superconductivity with magnetic field shows that a novel thermal metal-to-insulator transition occurs upon going
from the superconducting state to the field-induced normal state.
\end{abstract}
\begin{keyword}
Thermal transport \sep Metal-insulator transition \sep La$_{2-x}$Sr$_x$CuO$_{4}$
\end{keyword}

\end{frontmatter}

In underdoped La$_{2-x}$Sr$_x$CuO$_{4+\delta}$ (LSCO), resistivity measurements have revealed the field-induced
normal state to be a charge insulator \cite{Ando}.
On the other hand, the superconducting state of underdoped LSCO is a thermal metal,
in the sense that there is a clear $T$-linear contribution to the thermal conductivity at $T \rightarrow 0$
\cite{Sutherland,Takeya}.
Given that in all superconductors investigated to date (including cuprates) heat transport at $T \rightarrow 0$
is always seen to increase as one goes from the superconducting state to the field-induced normal state,
these two observations point to a violation of the Wiedemann-Franz law in underdoped cuprates.

In this article, we show the natural assumption that heat conduction will increase upon going
from the superconducting state to the field-induced normal state to be incorrect in underdoped LSCO.
Indeed, in the $T \rightarrow 0$
limit the thermal conductivity {\it decreases} in the vortex state and the residual linear term drops to
a value below our resolution limit in the field-induced normal state.
This result argues strongly for a thermally insulating normal state and reveals a novel thermal metal-to-insulator
transition.

Measurements of the thermal conductivity ($\kappa $) were performed down to 40~mK in fields up to
13~T on single crystals of  La$_{2-x}$Sr$_x$CuO$_{4+\delta }$ with $x=0$ (not superconducting)
and 0.06 ($T_\mathrm{c}$ = 5.5~K). Additional sample and measurement details are provided
elsewhere \cite{Hawthorn}.

\begin{figure}[t]
\centering
\resizebox{\columnwidth}{!}{\includegraphics[angle=-90]{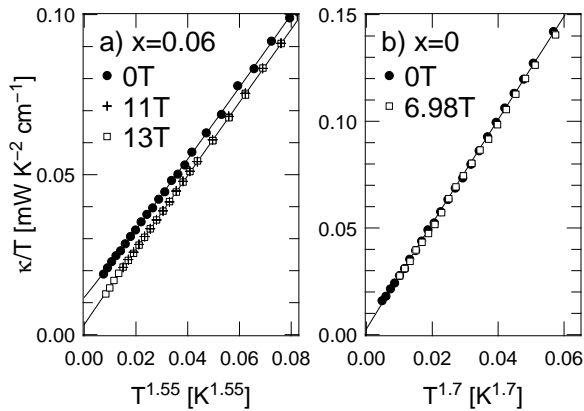}}
\caption{\label{fig:kappa}
$\kappa/T $ vs. $T^{\alpha -1}$ for La$_{2-x}$Sr$_x$CuO$_{4+\delta}$ with $x$ as shown.  The lines are fits to Eq.~\ref{eq:kappafit}.}
\end{figure}

In Fig.~\ref{fig:kappa} the thermal conductivity is plotted as $\kappa/T$ vs. $T^{\alpha -1}$,
where $\alpha $ is a free fitting parameter.
This type of plot is used to separate the electronic
($\kappa_\mathrm{el}$)
and lattice
($\kappa_\mathrm{ph}$)
contributions to $\kappa $ by making use of their different
power-law temperature dependences in the $T~\rightarrow~0$ limit.
In the limit $T \to 0$, $\kappa_\mathrm{el}$ is linear in $T$ for a $d$-wave superconductor on
account of nodal quasiparticle excitations \cite{Durst}.
Quite generally, a linear contribution to $\kappa $ at $T\to 0$
is direct evidence for fermionic excitations.
The phonon contribution can be modelled as
$\kappa_\mathrm{ph}~\propto~T^\alpha $ for phonons limited to scattering from the boundaries of the sample (see Ref.~\cite{Sutherland}).
Thus, $\kappa_\mathrm{el}$ and $\kappa_\mathrm{ph}$ can be separated by fitting the
data at low-temperatures to
\begin{equation}
\centering
\frac{\kappa }{T} = \frac{\kappa_0}{T} + BT^{\alpha -1} .
\label{eq:kappafit}
\end{equation}
The two distinct contributions are identified in Fig.~\ref{fig:kappa} as the intercept and slope of the curves, respectively, when plotting the data as
$\kappa /T$ vs. $T^{\alpha -1}$.

Having described our analysis, several observations can be made.
Firstly, in zero field (solid circles) the $x=0.06$ data reproduces the results of Refs.~\cite{Takeya} and \cite{Sutherland}
whereby a finite residual linear term in $\kappa(T)$ is observed at all superconducting dopings.
This proves the existence of delocalized zero-energy quasiparticles in the superconducting state.
In other words, the $d$-wave superconducting state is a thermal metal (see also \cite{Sutherland}).
In the $x=0$ sample, however, $\kappa_0/T$ becomes extremely small ($3~\mu$W~K$^{-2}$~cm$^{-1}$).
Now, even though
Eq.~\ref{eq:kappafit} provides
a good description of the data,
all the way up to 0.4~K,
the fact that $\kappa_0/T$ is 5 times smaller than the value of $\kappa/T$ at the lowest data point (40~mK)
means that one has to view the extrapolated value with caution.
The conservative position is to assume that the parent compound $x=0$ is a heat insulator
as well as a charge insulator, and regard this minute linear term of
$3~\mu$W~K$^{-2}$~cm$^{-1}$ as the resolution limit
of our technique, and treat the $x=0$ data as our reference (for an insulating state in LSCO).
By contrast, the linear term in the
$x=0.06$ sample (at 0T) of $12~\mu$W~K$^{-2}$~cm$^{-1}$
is clearly above the reference limit (by a factor 4)
and is thus unambiguously a thermal metal.

\begin{figure}[t]
\centering
\resizebox{\columnwidth}{!}{\includegraphics{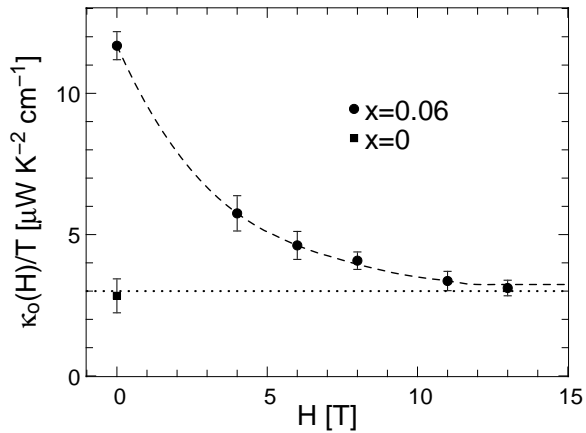}}
\caption{\label{fig:kvsH}$\kappa_0(H)/T$ vs $H$.   $\kappa_0/T$ is also
 shown for $x=0$ at zero field.  The dotted line represents the estimated resolution of our experiment
 (see text). The error bars are statistical errors in the fitted values of $\kappa_0/T$.}
\end{figure}

This brings us to the principal observation of this article: $\kappa $ {\it decreases with
increasing field for the x=0.06 sample}, as shown in Fig.~\ref{fig:kvsH} by the field evolution
of $\kappa_0/T$. This decreasing field dependence is in stark contrast to the increase in the
electronic heat conductivity in all other known superconductors at $T~\to~0$, including cuprates
at higher doping~\cite{Hawthorn,Chiao2}. Note that $\kappa$ is totally independent of magnetic
field in our reference sample ($x=0$). This shows that field dependence is a property of the
superconducting state. We can therefore use this criterion to establish that the
non-superconducting normal state is reached in the bulk by 11~T in sample $x=0.06$. Indeed, as
seen in Fig.~1a, a further increase of the field to 13~T causes no further change in $\kappa$.
This claim is supported by resistivity measurements on the same sample where the resistive onset
of superconductivity is entirely absent for fields of 12~T and above (down to
40~mK)~\cite{Hawthorn}. We take this as an additional indication that the field-induced
(non-superconducting) normal state has been reached by 13~T at $x=0.06$ (in the bulk).  Moreover,
as seen in Fig.~\ref{fig:kvsH}, $\kappa_0/T$ drops by a factor 4 from $H=0$ to $H=13$~T, where it
reaches a value equal to that of the reference sample, namely $\kappa_0/T =
3~\mu$W~K$^{-2}$~cm$^{-1}$. We thus conclude that the field-induced normal state in underdoped
LSCO is a thermal insulator. {\em This implies the existence of an unprecedented kind of thermal
metal-to-insulator transition.} The superconducting state is a thermal metal by virtue of its
delocalized nodal quasiparticles, while the field-induced normal state in the same sample is a
thermal insulator, with either no fermionic excitations or localized fermionic excitations.

In summary, we have observed in underdoped LSCO a decrease in thermal conductivity with magnetic field
upon going from the superconducting state to the field-induced normal state.
 We show that this result is due to a novel
thermal metal-to-insulator transition.


This work was supported by the Canadian Institute for Advanced Research and NSERC of Canada. We
would like to thank H. Dabkowska, B.D. Gaulin and G. Luke for use of their image furnace.


\end{document}